\documentclass[prl,twocolumn,showpacs]{revtex4-1}
\usepackage{amsmath}
\usepackage{dsfont}
\usepackage{bm}
\usepackage{graphicx}
\usepackage[applemac]{inputenc}
\usepackage[T1]{fontenc}
\usepackage{amssymb}
\usepackage{siunitx}
\usepackage{datetime}
\usepackage{color}
\usepackage{times}

\begin{document}
\newcommand{\freiburg}{Physikalisches Institut, Albert-Ludwigs-Universit\"{a}t Freiburg, Hermann-Herder-Stra{\ss}e 3, D-79104, Freiburg, Germany}
\title{Interaction effects on dynamical localization in driven helium}
\author{Felix J\"order}
\author{Klaus Zimmermann}
\author{Alberto Rodriguez}
\author{Andreas Buchleitner}
\email[]{a.buchleitner@physik.uni-freiburg.de}
\affiliation{\freiburg}
\date{$Revision: 226 $, compiled \today, \currenttime}
\begin{abstract}
Dynamical localization prevents driven atomic systems from fast fragmentation by hampering the excitation process. 
We present numerical simulations within a collinear model of microwave-driven helium Rydberg atoms 
and prove that dynamical localization survives the impact of electron-electron interaction, even for doubly excited states in the presence of fast autoionization. 
We conclude that the effect of electron-electron repulsion on localization can be described by an appropriate rescaling of the atomic level density and of the external field with the strength of the interaction. 
\end{abstract}
\maketitle
\begin{figure}
\centering
\includegraphics[width=0.9\columnwidth]{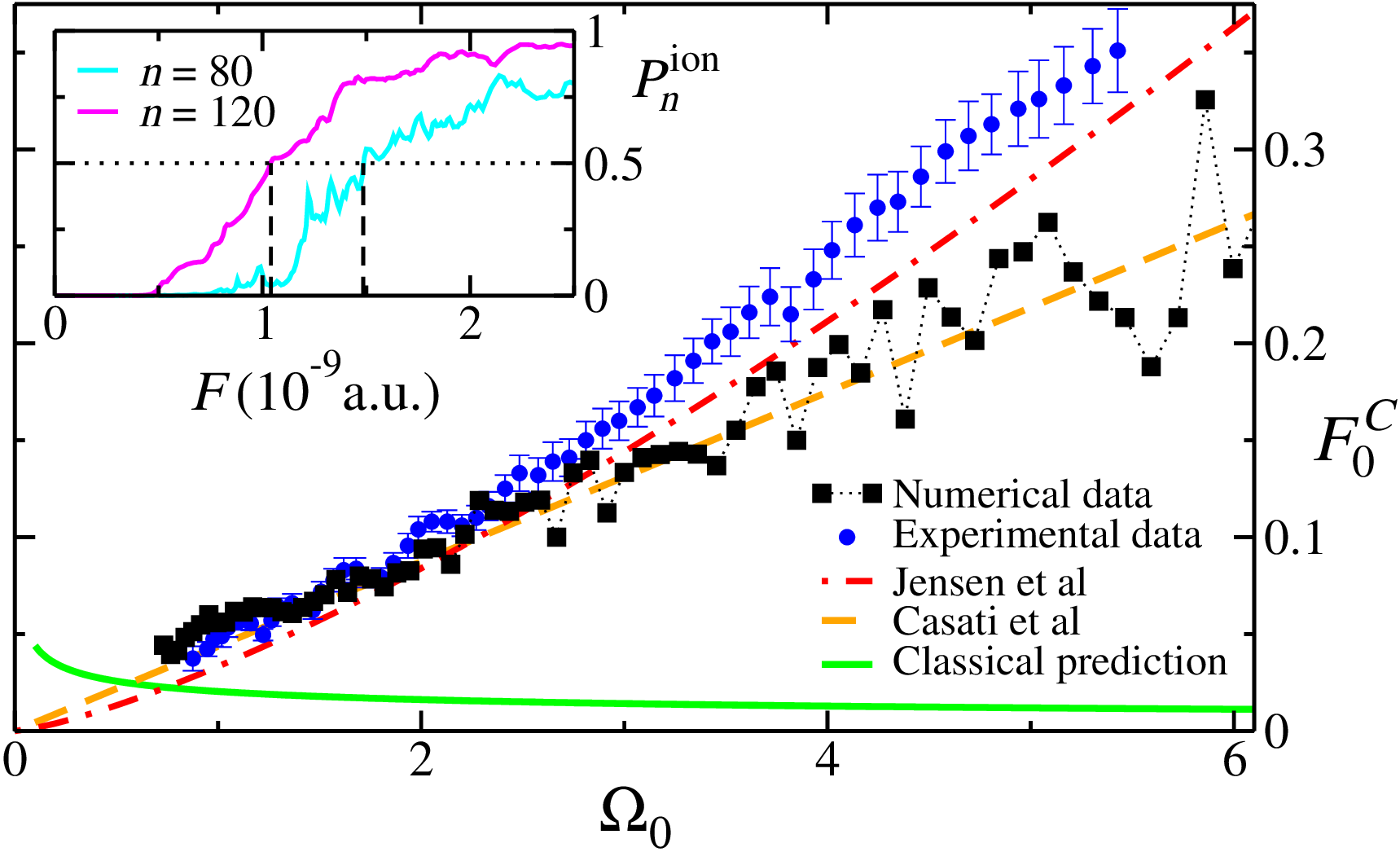}
\caption{(color online) Scaled critical fields $F^C_0=n^4F^C$ vs scaled frequencies $\Omega_0=n^3\omega$ ($\omega=\SI{2.667e-6}{{a.u.}}$), for the second Rydberg series of $Zee$ helium: $N=2$ and $n\in[65,130]$. Our numerical data for helium are indicated by black squares. Experimental results for $Sr$ are reproduced from Ref.~\cite{PRL_Maeda_2004}. 
The dashed line marks the prediction for 1D hydrogen given by Casati \emph{et al.}~\cite{IEEE_Casati_88}. The dashed-dotted line indicates the result given by Jensen \emph{et al.}~\cite{PRL_Jensen_89}. The solid line corresponds to the classical prediction \cite{IEEE_Casati_88}. The inset shows ionization yield $P_n^\text{ion}$ vs field strength $F$ for two states $n=80,120$. Vertical lines mark the required critical fields to yield $50\%$ ionization after a driving time of $\SI{20}{ns}$.}
\label{fig:Fcrit_01} 
\end{figure}

Anderson localization implies the exponential decay of wavefunctions in configuration space due to a disordered potential landscape \cite{PR_anderson_58},
and is well understood as a single-particle effect \cite{BilJZB08,RoaDFF08,JenBMC12,KonMZD11,HuSPS08,SchBFS07}. 
\emph{Dynamical localization}, its analogue emerging in periodically driven quantum systems \cite{PRL_Fishman_82}, is best known in connection with the quantum kicked rotor, which 
has enjoyed quite an experimental success in the field of cold atoms \cite{bharucha1999,RinSGD00,ChaLGD08,LemLDS10,MatCSG12,LopCLD13,ArcSFG04,BehRAS06}. 
There, a pseudo-randomness generated by the kicked dynamics leads to an exponential suppression of transitions to high momentum states. 
In the context of electromagnetically driven atomic systems, the excitation energy takes the role of the localized observable: excitation processes corresponding to an energy exchange of $E=k\omega$ exhibit an exponential decrease $\exp{(-|k|/\xi)}$, where $|k|$ is the net number of exchanged photons and $\xi$ is the localization length 
\cite{PRA_Casati_1987,PRL_Bayfield_89,PRL_Galvez_1988,PRL_Jensen_89,PRL_Abu_2009,IEEE_Casati_88}.
The suppression of the excitation process translates into highly enhanced values of the critical field strength $F^{C}$ needed to ionize the atoms 
with respect to the classical prediction. 
This effect has been measured experimentally \cite{PRL_Bayfield_89,PRL_Galvez_1988,ArnBMW91,PRL_Maeda_2004} and is supported by numerical simulations for hydrogen and alkali atoms \cite{IEEE_Casati_88,Krub01}, and also by our results on collinear helium, which show a significant agreement with the experimental data (cp.~Fig.~\ref{fig:Fcrit_01}).

In this letter we approach the problem on how localization is affected by particle-particle interactions. This is a broad-interest question which is currently being intensively investigated in the context of metal-insulator transitions and 
many-body Anderson localization \cite{DeiZRD10,PolPST09,AniKPF07,AleAS10,SanL10}.
For two particles in one dimension, 
short-range interactions can weaken localization leading to an enhancement of the two-particle localization length 
\cite{Dor90,PRL_shepelyansky_1994,EPL_Imry_1995,PorS95,PRB_Evangelou_1996,PRL_vonOppen_1996,EPL_shepelyansky_1997,PonS97,PhysStatSol_Roemer_1999,PRB_Song_1999,JETPL_Krimer_2011}.
In the case of the quantum kicked rotor,
which can be mapped onto the standard Anderson problem \cite{PRL_Fishman_82},
interactions in the form of nonlinearities are known to induce delocalization \cite{PRL_shepelyansky_1993,EPL_Gligoric_2011}.
While the driven one-electron atom in turn maps onto the quantum kicked rotor \cite{PRA_Casati_1987},
the extension of this mapping to the many-particle case is non-trivial.
Thus, the study of dynamical localization in the presence of a long-range Coulomb interaction 
considerably expands the realm of localization phenomena within the area of light matter interaction, where the interelectronic repulsion is essential, e.g., to understand the correlations observed in laser-driven atomic ionization \cite{BecLHE12,BilGWE13,ZhoHLL12,CamFKS12,WalSDA94,WebGWU00,MosFLC02}.
The helium atom is a fundamental and experimentally accessible system which can be used to shed some light onto this subject.

\paragraph{Field-free helium.---}
We consider two distinct configurations of collinear helium, namely $Zee$ and $eZe$ helium, characterized by whether the electrons are located on the same side or on opposite sides of the nucleus, which is fixed at the origin. 
The field-free Hamiltonian in atomic units reads
\begin{equation}
  H_0=\frac{1}{2}\left(p_1^2+p_2^2\right)-\frac{Z}{r_1}-\frac{Z}{r_2}+\frac{\gamma}{r_{12}},
  \label{equH0}
\end{equation}
for $Z=2$, where $r_1,r_2>0$ denote the radial coordinates of the electrons, and $r_{12}=|r_1\mp r_2|$ for $Zee(-)$ and $eZe(+)$ helium. 
The radial momentum operators are $p_j\equiv-i\partial_{r_j}$.
The parameter $\gamma$, with a physical value $\gamma=1$, tunes the strength of the electronic interaction.
The classical dynamics underlying the two configurations are rather distinct and exhibit complementary features
of the three-dimensional problem.
While the $eZe$ phase space is fully chaotic \cite{JPhysB_Ezra_1991}, leading to fast fragmentation,
long-lived 'frozen planet states' exist for $Zee$ helium --- which, although with shorter lifetimes, persist in two and three dimensions \cite{PRL_Richter_1990,LopAO96,SchB98,EPL_Madronero_2005}.
Furthermore, both collinear configurations are stable against small transverse displacement in three dimensions \cite{PRA_Richter_93,RevModPhys_Tanner_2000}.

The interaction term $\gamma/r_{12}$ in $H_0$ induces a coupling of all doubly excited states to the underlying single-ionization continua, and thus turns them into resonances with finite decay (autoionization) rates.
The energies $E_j$, decay rates $\Gamma_j$ and wavefunctions $\phi_j$ associated to these resonances can be calculated from the eigenvalues and eigenstates of a non-hermitian Hamiltonian obtained by the method of complex rotation \cite{ComMathPhys_Balslev_1971}, using a basis of Sturmian functions \cite{SchB98,SchB03}.
In order to treat exactly all the Coulomb terms in Eq.~\eqref{equH0} we work with a regularized $H_0$,
multiplied by $r_1r_2r_{12}$. 

The discrete spectrum of collinear helium can be characterized by quantum numbers $(N,n)$, and it is organized in 
Rydberg series which converge to single ionization thresholds $E_N=-Z^2N^{-2}/2$ \cite{RevModPhys_Tanner_2000}.
In a typical Rydberg state ($N=2,n\sim100$) we find autoionization rates for $Zee$-helium corresponding to long lifetimes $\tau_\text{typ}=\Gamma_\text{typ}^{-1}\sim\SI{300}{\micro\second}$, in accordance with the classical stability of this configuration. For $eZe$ we obtain two sets of resonances, even and odd with respect to the exchange of $r_1$ and $r_2$, 
with  $\tau_\text{typ}\sim\SI{5}{\nano\second}$ and $\SI{500}{\nano\second}$, respectively. 

\paragraph{Driven helium.---}
We take into account a dipole coupling to an external classical electromagnetic field of strength $F$ and frequency $\omega$. The Hamiltonian in velocity gauge reads $H=H_0 + H_F$, where
\begin{equation}
H_F=\frac{F}{\omega}(p_1\pm p_2)\sin(\omega t)=V_F \sin(\omega t),
\end{equation}
for $Zee(+)$ and $eZe(-)$.
Due to the periodicity of $H_F$, we may use Floquet theory to solve the time-dependent Schr\"odinger equation for the complex rotated Hamiltonian. 
The elementary solutions have the form 
\begin{equation}
\psi_\epsilon(t)=e^{-i\epsilon t}\sum_{k=-\infty}^\infty e^{-i k \omega t}\psi_\epsilon^k,
\label{equFloq1}
\end{equation}
where the Floquet components $\psi^k_\epsilon$ and quasienergies $\epsilon$ obey the eigenvalue equation
\begin{equation}
   (H_0-k\omega)\psi_\epsilon^{k} + \frac{1}{2i}V_F\left(\psi_\epsilon^{k+1} - \psi_\epsilon^{k-1}\right)=\epsilon\,\psi_\epsilon^{k}\,.
\label{equFloq2}
\end{equation}
The Floquet index $k$ can be effectively related to the number of photons exchanged between atom and field \cite{Shi65}. 
Given a reference state with energy $E_0$, we expand the components $\psi_\epsilon^{k}$ 
in terms of field-free atomic eigenstates lying inside an energy interval $\Delta E$, centered at $E_k=E_0+k\omega$. 
The tolerance $\Delta E$ then corresponds to the maximum allowed detuning of any $k$-photon transition.
Numerically, we detect convergence of the results by their invariance under a further increase of $\Delta E$, which is always the case for $\Delta E<10\,\omega$.
Using this method, we reduce the dimensionality of our Floquet eigenvalue problem,
which enables us to consider processes of very high order ($>100$) in $k$.

The field-induced transition probability between atomic levels $\phi_i$ and $\phi_j$ after an interaction time $T$ with the field, is given by
$P_{i\rightarrow j}(T)=\left|\langle\phi_j|U(T)|\phi_i\rangle\right|^2$, 
where the time evolution operator $U(T)$ is resolved in terms of the solutions in Eq.~\eqref{equFloq1}. 
The ionization probability of the initial atomic state $\phi_i$ is then obtained as
$P_i^{\text{ion}}(T)=1-\sum_jP_{i\rightarrow j}(T)$. 

For comparison with experimental data and theoretical predictions, we calculate ionization yields and critical fields $F^C$ of $Zee$ Rydberg states of the second series, $N=2$,
for the same parameters as those experimentally considered in Ref.~\cite{PRL_Maeda_2004}. 
Due to the low excitation of the inner electron and the slow autoionization, we expect no crucial differences in comparison to driven hydrogen or alkali atoms. 
We use the scaled variables 
$F_0^C=n^4F^C$, $\Omega_0=n^3\omega$,  
to display the results: $F_0^C$ is the critical field in units of the average Coulomb field in a hydrogen atom, and $\Omega_0$ is the frequency in units of the hydrogen level spacing. As shown in Fig.~\ref{fig:Fcrit_01}, our calculations for $Zee$ helium ---in spite of the one-dimensional nature of the model--- agree with the experimental results of Ref.~\cite{PRL_Maeda_2004} in the regime $\Omega_0\leqslant 2.5$ \cite{footnoteDeviations}.
Our data is also consistent with the localization-based prediction of Ref.~\cite{IEEE_Casati_88}.

\paragraph{Influence of electron-electron interaction on dynamical localization.---}
The energy of a doubly excited state can be cast into an effective Rydberg form via  
\begin{equation}
 E=-\frac{Z^2}{2N^2}-\frac{Z_\text{eff}^2}{2n_\text{eff}^2},
  \label{equQDef1}
\end{equation}
where $Z_\text{eff}=Z-\gamma$ and $\gamma\in[0,1]$.
The effective quantum number $n_\text{eff}$ encodes the influence of the electron-electron interaction on the spectrum.
For the $Zee$ configuration we found that 
$n_\text{eff}=n+\delta_N$, given by a so-called quantum defect $\delta_N$, which is only determined by the series index $N$. The whole spectrum of $Zee$ helium is described very  accurately by $\delta_N$ and Eq.~\eqref{equQDef1} \cite{Note:qdefect}.
Following this picture, we argue that the outer electron sees a mean level spacing $\Delta=n_\text{eff}^{-3}Z_{\text{eff}}^2$ and experiences a Coulomb field $f=n_\text{eff}^{-4} Z_{\text{eff}}^3$. The critical ionization fields $F^C$ and the driving frequency $\omega_\gamma$ should then be measured with respect to these fundamental quantities. The appropriately scaled variables in this case are
\begin{equation}
 \widetilde{F}^C_0\equiv\frac{F^C}{f}=n_\text{eff}^4 Z_{\text{eff}}^{-3}F^C, \quad 
 \widetilde{\Omega}_0 \equiv \frac{\omega_\gamma}{\Delta}=n_\text{eff}^{3} Z_\text{eff}^{-2}\omega_\gamma.
\end{equation}
The driving frequency is changed with $\gamma$ as $\omega_\gamma=Z_{\text{eff}}^2\,\omega_1$, for $\omega_1=\SI{1.508e-6}{}\text{ a.u.}$, 
which is one of the frequencies employed in the experiments by Koch \emph{et al.} \cite{PRL_Galvez_1988}.
This scaling ensures that for a given value of $\widetilde{\Omega}_0=n_\text{eff}^{3}\omega_1$ the number of photons needed to ionize the outer electron is the same for all $\gamma$.
This latter condition is crucial to compare the occurrence of localization for different interaction strengths. 

\begin{figure}
\centering
\includegraphics[width=0.9\columnwidth]{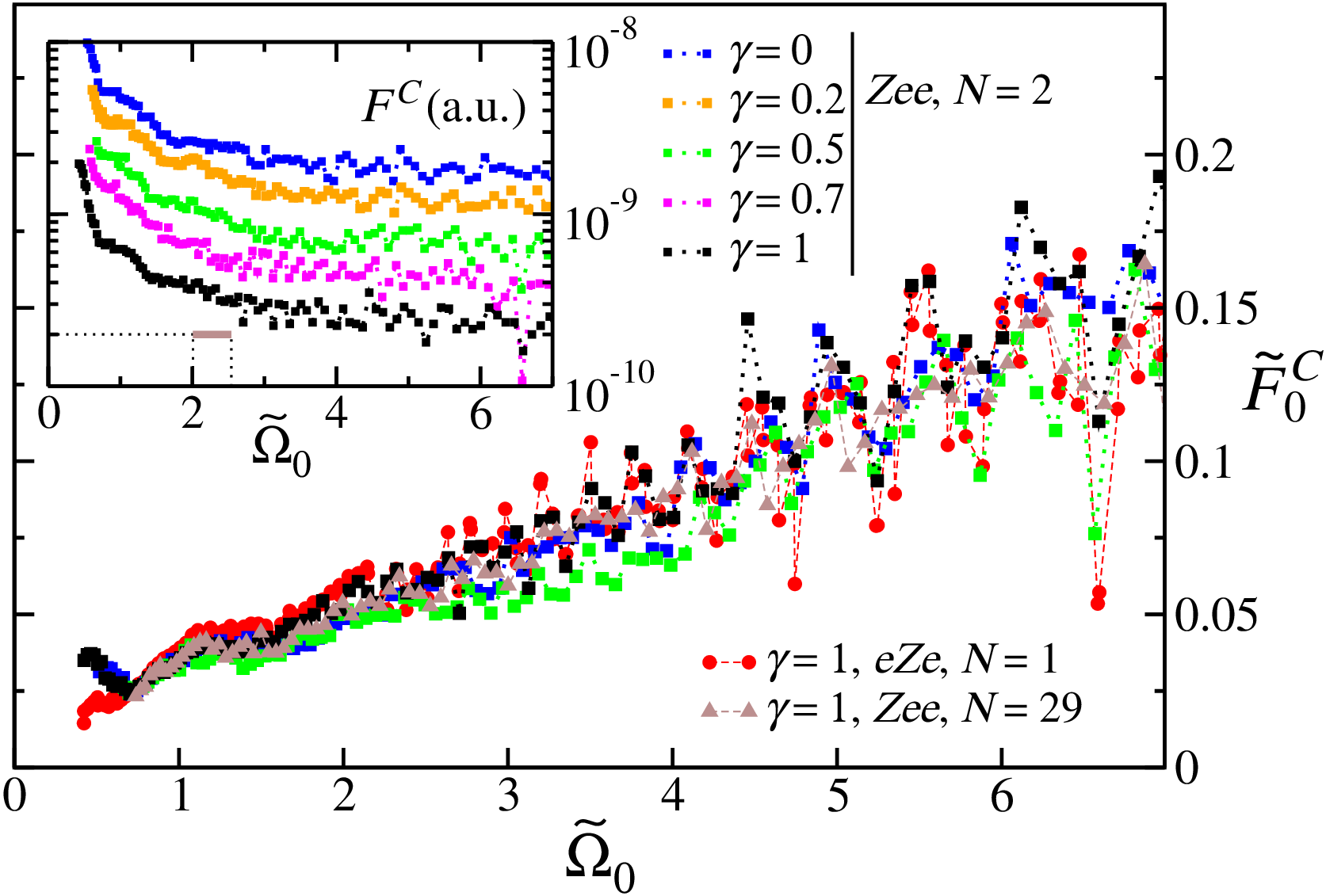}
\caption{(color online) Scaled critical fields $\widetilde{F}^C_0$ vs scaled frequencies $\widetilde{\Omega}_0$ for $20\%$ ionization after a driving time of $\SI{100}{\nano\second}$, for different series $N$ and values of the interaction strength $\gamma$. The shown range of $\widetilde{\Omega}_0$ corresponds to $n_\text{eff}\in[65, 170]$, and the number of absorbed photons needed for ionization ranges from 80 to 12. For $N=2$ the quantum defect is negligible whereas $\delta_{N=29}^{Zee}\simeq 14$. The inset shows the absolute critical fields $F^C$ vs $\widetilde{\Omega}_0$ for $N=2$ and $Zee$ helium. The horizontal bar in the inset marks the parameter region considered for the simulations of Fig.~\ref{fig:population}.}
\label{fig:Fcrit_inset_02}
\end{figure}

We first calculate the critical fields $F^C=F^{20\%}$ yielding  $20\%$ ionization after a driving time of $\SI{100}{\nano\second}$ ($\sim10^3 Z_\text{eff}^2$ field periods), of states in the second Rydberg series of $Zee$ helium ($N=2$) for different values of $\gamma$.
The obtained $F^C$ are shown in the inset of Fig.~\ref{fig:Fcrit_inset_02}. For a fixed $\widetilde{\Omega}_0$, the critical field decreases by almost one order of magnitude with increasing $\gamma$. Therefore, the ionization process is strongly enhanced due to the electron-electron interaction, which in turn reflects a weakening of localization.
Nevertheless, as seen in Fig.~\ref{fig:Fcrit_inset_02}, the values of the scaled field $\widetilde{F}^C_0$ remarkably coincide for all $\gamma$ up to fluctuations, indicating that the absolute critical fields $F^C$ are proportional to $Z_\text{eff}^{3}$. 
We found that this result carries over to the first Rydberg series $(N=1)$ of $eZe$ helium, also shown in Fig.~\ref{fig:Fcrit_inset_02}.
We also considered the high series $N=29$ for $Zee$ helium, where the electron-electron interaction effects are very strong, as indicated by 
the value of the quantum defect $\delta_{N=29}^{Zee}\simeq 14$. As shown in Fig.~\ref{fig:Fcrit_inset_02}, the scaled critical fields again coincide with the data for the low-lying series. 

The collapse of the critical field curves suggests that the influence of the inner electron on the ionization dynamics can be understood from the appropriate rescaling of the level spacing and  the elementary Coulomb field for the outer electron.
In the case of doubly excited $eZe$ states, however, fast autoionization
makes the definition of critical fields meaningless. In order to show that dynamical localization 
is present, we monitor directly the field-induced transitions $P_{i\rightarrow j}(T)$ between atomic levels.
The population redistribution of a driven initial state $\phi_i$ is then visualized in energy space as a function of 
the excitation energy $\Delta E=E_j-E_i$, measured in multiples $k\in\mathbb Z$ of the photon energy. 
The resulting populations $P(k)$ can thus be correlated to the net number of photons emitted and
absorbed.
\begin{figure}
\centering
\includegraphics[width=.95\columnwidth]{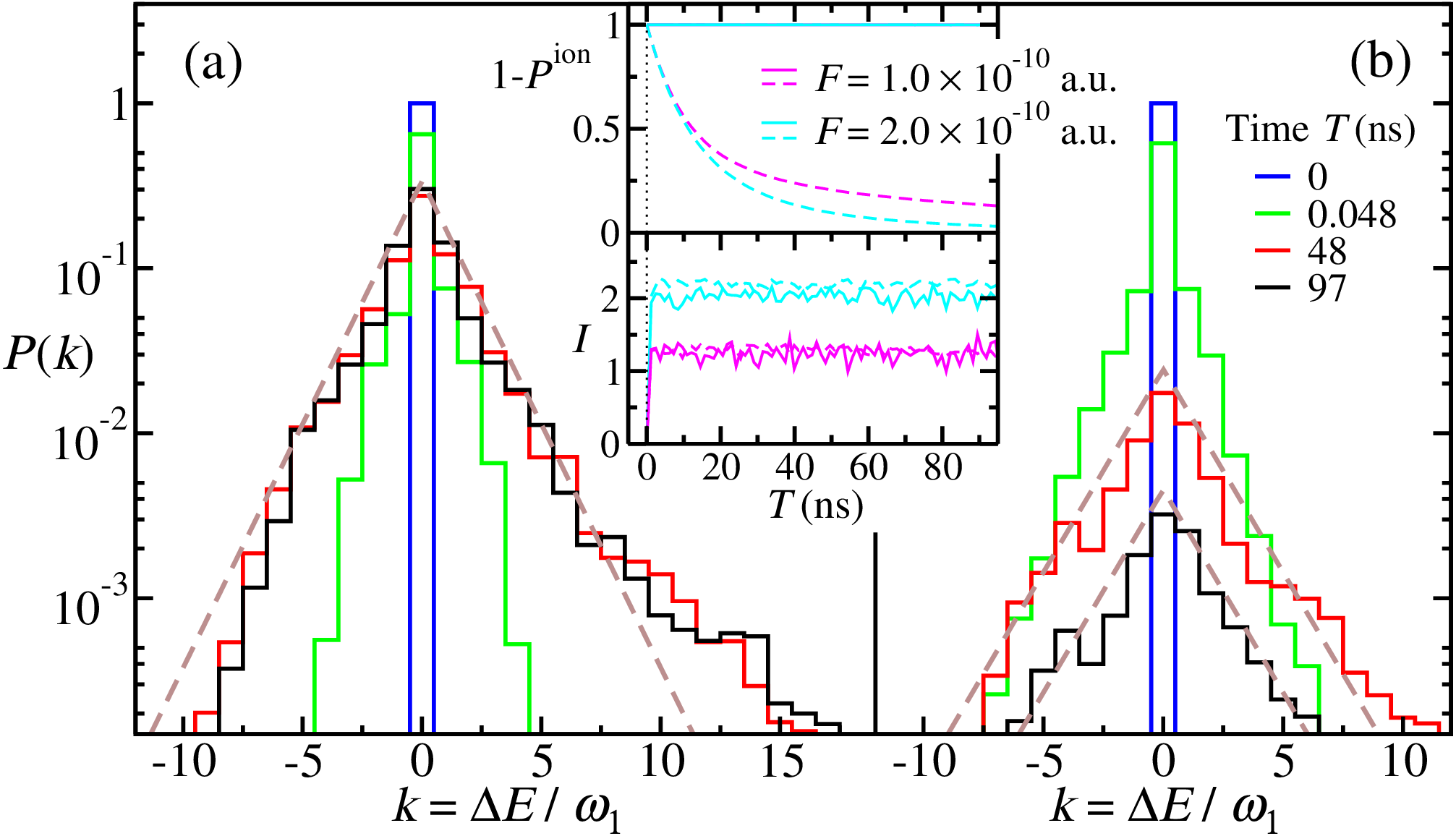}
\caption{(color online) Histogram of transition probabilities $P(k)$ vs excitation energy in units of $\omega_1\,(\gamma=1)$, averaged 
over initial states $n\in[110,119]$ of the $N=2$ Rydberg series of (a) $Zee$ and (b) $eZe$ helium, 
for a field strength of $F=\SI{2e-10}{\text{a.u.}}$ (see corresponding $\widetilde{\Omega}_0$ range in the inset of Fig.~\ref{fig:Fcrit_inset_02}), after a driving time $T$.
Thick dashed lines highlight the exponential decay according to the localization length obtained from the analysis of the Shannon entropy, $I$. 
The upper and lower insets show respectively the probability of the outer electron to remain bound to the atom, $1-P^\text{ion}$, and $I$ of the normalized distributions vs $T$ (dashed lines: $eZe$, solid lines: $Zee$), averaged over the same states. 
Fast autoionization of the $eZe$ states is also observed for smaller fields of $F=\SI{1e-10}{\text{a.u.}}$,
due to average autoionization rates $\bar\Gamma^{eZe}_\text{even}=\SI{3.3e-9}{{a.u.}}$ and $\bar\Gamma^{eZe}_\text{odd}=\SI{3.1e-11}{{a.u.}}$ ($\bar\Gamma^{Zee}=\SI{6.9e-14}{{a.u.}}$).
}
\label{fig:population}
\end{figure}
Figure \ref{fig:population} shows these distributions for the second series of $Zee$ and $eZe$,
averaged over initial states $n\in[110,119]$, for several driving 
times. 
The chosen field strength, $F=\SI{2e-10}{\text{a.u.}}$, 
is strong enough to observe a considerable spreading, but still 
does not lead to ionization of the $Zee$ initial states (see the bar in the inset of Fig.~\ref{fig:Fcrit_inset_02}).
In the $Zee$ case the distribution freezes completely after about $\SI{10}{\nano\second}$ and approaches
an exponentially localized shape.
For the $eZe$ states, the distribution also localizes exponentially in $k$, but its norm decreases with time due to fast autoionization, as seen 
in the upper inset of Fig.~\ref{fig:population}. 
We characterize the width of the normalized distribution $\hat P(k)$ via the Shannon entropy, 
$I=-\sum_k\hat P(k)\log{\hat P(k)}$.
As depicted in the lower inset of Fig.~\ref{fig:population}, $I$ increases rapidly for short times and then
fluctuates around some saturated value. 
Hence, the field-induced transport on the energy axis freezes, and we conclude that localization is still present despite any loss of
norm due to autoionization. 

In order to carry out a quantitative analysis of the localization behavior, 
we estimate the localization length $\xi$ of the distributions $P(k)\sim\exp{(-|k|/\xi)}$ from 
the limiting value of $I$, averaged over times between $48$ and $\SI{97}{\nano\second}$ \cite{Note:shannon}.
For the range of initial states considered, $\xi$ seems to be roughly independent of $n$, apart from 
fluctuations caused by the local detuning of the field-induced transitions.  
The estimate of the localization length and its uncertainty are obtained from the average over initial states $n\in[110,119]$. 
We studied the dependence of $\xi$ on the field strength $F$ for the second Rydberg series of both helium configurations, as well as its dependence on the interaction
strength $\gamma$ for $Zee$ helium. 
\begin{figure}
\centerline{\includegraphics[width=0.9\columnwidth]{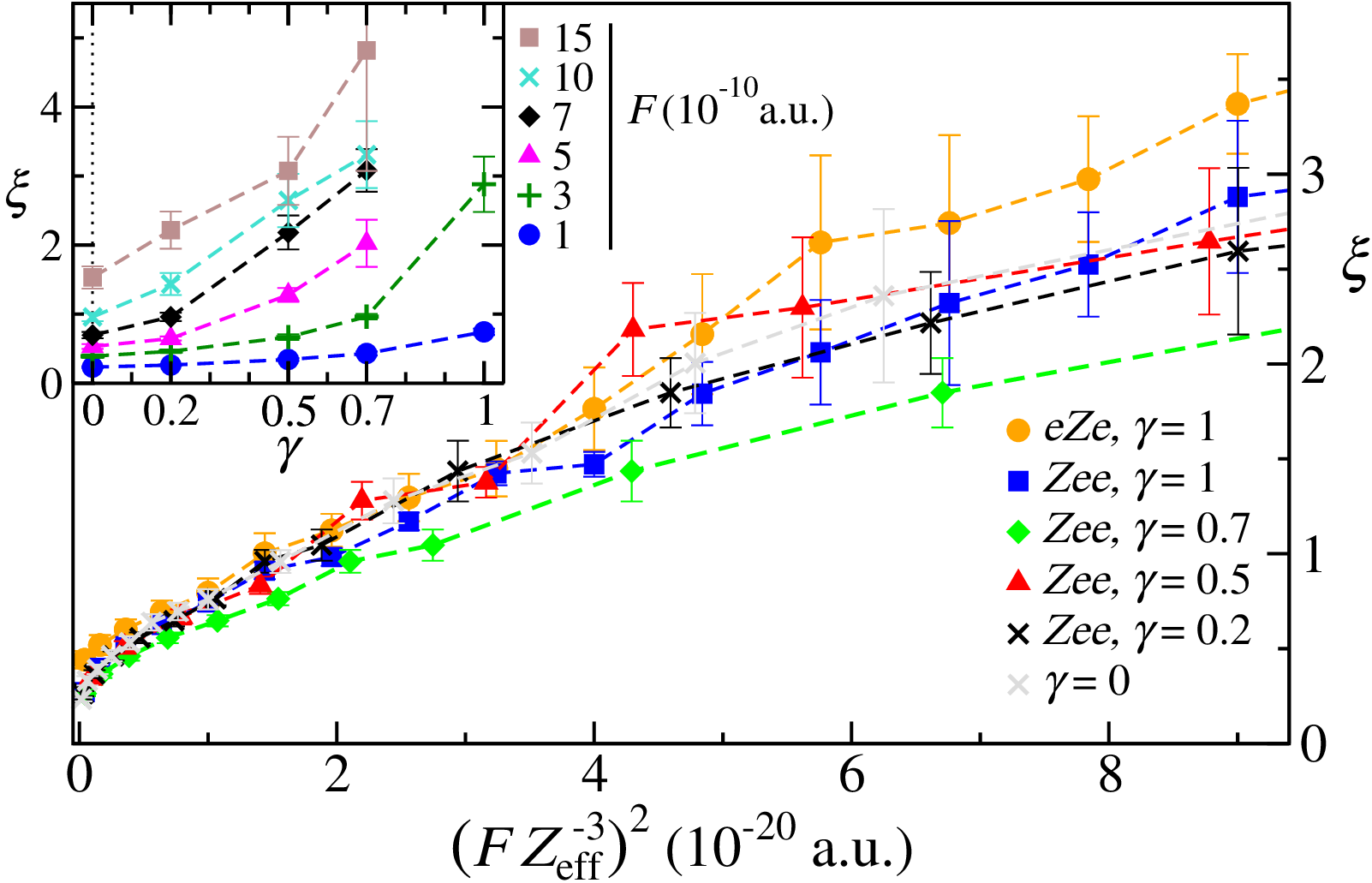}}
\caption{(color online) Localization length $\xi$ for the $N=2$ Rydberg series of  $Zee$ and $eZe$ helium, obtained from the Shannon information $I$ of the distributions $P(k)$ [cp.~Fig.~\ref{fig:population}], as a function of $(F Z_\text{eff}^{-3})^2$. An average of $I$ for $T\in[48,97]\SI{}{\nano\second}$ and over initial states $n\in[110,119]$ has been considered.
Error bars indicate one standard deviation. 
Higher $F$ implies larger fluctuations of $\xi$ as a function of $n$, which results in larger uncertainties.
The inset shows $\xi$ vs interaction strength $\gamma$ for $Zee$ states and several field intensities $F$.}
\label{fig:chi_03}
\end{figure}
As shown in the inset of Fig.~\ref{fig:chi_03}, for a fixed field strength $\xi$ increases 
with $\gamma$, once more demonstrating the enhancement of the excitation process due to the interaction. For a fixed $\gamma$ the localization length also 
grows with the field strength $F$.
As discussed above, the critical fields scale with $Z_\text{eff}^3$. 
Therefore, in order to treat different interaction strengths on an equal footing, we should rescale $F$.  
As shown in Fig.~\ref{fig:chi_03} the values of $\xi$ as a function of $F Z^{-3}_\text{eff}$ overlap within their $95\%$ confidence interval for all values of $\gamma$.
Therefore, the influence of the electron-electron interaction on localization can be described by taking into account an appropriate rescaling of the field strength with the effective charge.
Additionally, we observe that the estimated localization lengths for $FZ_\text{eff}^{-3}\gtrsim10^{-10}\,\text{a.u.}$ are seemingly compatible with a quadratic scaling law, 
$\xi\sim (F\,Z_\text{eff}^{-3})^2$ , as found for driven hydrogen \cite{PRA_Casati_1987}.
Motivated by the scaling law for the two-particle localization length 
put forward by Shepelyansky \cite{PRL_shepelyansky_1994} and Imry \cite{EPL_Imry_1995}, we have analysed the enhancement of $\xi$ in the form 
$\xi(\gamma\neq0)\sim \xi(\gamma=0)^\kappa$, finding a $\gamma$-dependent $\kappa$ \cite{PonS97} that increases from $\kappa\simeq1$ for $\gamma\lesssim0.2$ to $\kappa\simeq2$ for $\gamma=1$. Values $1<\kappa<2$ have also been observed for two particles with on-site interaction in configuration space \cite{JETPL_Krimer_2011}.

In conclusion, using efficient numerical techniques we have verified the suppression of field-induced single-particle ionization in helium due to dynamical localization. 
We have shown that the effect of electron-electron interaction on localization can be described through the influence of the former on the atomic level density and an appropriate scaling of the external field with the effective charge.
This also holds in the presence of fast autoionization as a dominant competing process.
We emphasize that dynamical localization in the presence of autoionization is the equivalent of Anderson localization in the presence of absorption, 
which poses a challenge for the observation of light localization \cite{CheST10,SpeBAM13}.

We expect our results to hold even for comparable quantum numbers of both electrons.
While neighboring Rydberg series of the unperturbed $eZe$ configuration strongly mix for $N\gtrsim 20$,
strong mixing of odd and even states is induced by the driving field at lower values of $N$,
and does not affect the localization mechanism, as we have shown for $N=2$.
Therefore, we conjecture that localization
will persist also for $N>20$, what will have to be verified in future work.
Another future perspective is the characterization of the
fluctuations of the photoionization signal under parameter variations.
The statistics thereof is a sensitive indicator of the underlying transport mechanism \cite{JPhysA_wimberger_2001}
and might allow for a refined assessment of the fingerprints of electron-electron interactions.

We thank P.~Lugan for fruitful discussions and numerical support during the early stages of this project.
The authors gratefully acknowledge the Deutsche Forschungsgemeinschaft for financial support, 
and the computing time granted by the John von Neumann Institute for Computing, 
and provided on the supercomputer JUROPA at J\"ulich Supercomputing Centre.
Numerical calculations were also performed on the Black Forest grid (Freiburg).
\bibliographystyle{prsty}

\begin{thebibliography}{10}
\bibitem{footnoteDeviations}
 The disagreement observed for larger $\Omega_0$ in Fig.~\ref{fig:Fcrit_01} might be due to the participation of states with non-vanishing angular momentum in the excitation process in three dimensions.
\end{thebibliography}

\end{document}